\documentclass[reprint,amsmath,amssymb,aps,superscriptaddress]{revtex4-1} 
\usepackage[utf8]{inputenc}

\usepackage{amsmath}
\usepackage{euscript}
\usepackage{graphicx}
\usepackage{subcaption}
\usepackage{amsfonts}
\usepackage{float}
\usepackage[thinlines]{easytable}
\usepackage{multirow}

\usepackage{listings}
\usepackage{leftidx}

\usepackage{braket}

\usepackage{amsmath,amsthm,amssymb,amsfonts}
\usepackage{graphicx}
\usepackage[colorinlistoftodos]{todonotes}
\usepackage[colorlinks=true, allcolors=blue]{hyperref}
\usepackage{xpatch}
\xpatchcmd{\proof}{\itshape}{}{}

\begin{document}
\title{Digital quantum simulation of beam splitters and squeezing with IBM quantum computers}
\author{Paula Cordero Encinar}
  \author{Andrés Agustí}
\affiliation{Instituto de Física Fundamental, CSIC, Serrano 113-bis 28006 Madrid, Spain}
    \author{Carlos Sabín}
\affiliation{Departamento de Física Teórica, Universidad Autónoma de Madrid (UAM), 28049 Madrid, Spain}
\begin{abstract}
We present results on the  digital quantum simulations of beam-splitter and squeezing interactions. The bosonic hamiltonians are mapped to qubits and then digitalized in order to implement them in the IBM quantum devices. We use error mitigation and post-selection to achieve high-fidelity digital quantum simulations of single-mode and two-mode interactions, as evidenced --where possible-- by full tomography of the resulting states. We achieve fidelities  above 90 \% in the case of single-mode squeezing with low squeezing values and ranging from 60 \% to 90 \% for large squeezing and in the more complex two-mode interactions.
\end{abstract}
\maketitle

\section{Introduction}
Paradigmatic quantum-optical processes such as single-mode and two-mode squeezers and beam-splitter interactions among bosonic modes are now at the heart of quantum information processing and quantum computation with continuous variables \cite{gao,zhanggirvin,schoelkopfgirvin}, including the quantum simulation of molecular vibronic spectra \cite{pollo, polloexp} and the achievement of quantum supremacy with gaussian boson sampling \cite{supremacybs}. 

In parallel, qubit-based %--specially, superconducting-qubit ones--
quantum computers and simulators are rapidly transitioning from small-scale experiments to large networks \cite{boixo,naturemonty} in which quantum supremacy has also been claimed \cite{arutesup}. It is natural to look for links between these two alternate approaches to modern quantum technologies. Recently, one of us has proposed a recipe for digital quantum simulation of multimode bosonic hamiltonians \cite{sabindqs}. The idea is  to combine Trotter \cite{lloyd} and gate-decomposition techniques \cite{sommalallama} with boson-qubit mappings \cite{losalamos,sommathesis} in order to encode generic bosonic hamiltonians into a sequence of single-qubit and two-qubit gates. 

In this work, we apply these ideas to the digital quantum simulation in IBM quantum devices of single-mode and two-mode squeezing and beam-splitter interaction hamiltonians. We make use of error mitigation and post-selection techniques to achieve high-fidelity quantum simulations, which we demonstrate by performing --where possible-- full tomography of the final quantum states, or otherwise employing analytical approximations of the fidelity.  We achieve fidelities around 90 \% in the case of single-mode squeezing with low values of the squeezing parameter and ranging from 60 \% to 90 \% in the case of two-mode squeezing and two-mode beam-splitting, which require a larger number of two-qubit gates. Fidelity also decreases for large squeezing parameters in the single-mode case, which would require more qubits in order to simulate more photonic excitations.

The structure of the paper is the following. In Section II we thoroughly discuss the digitalization and quantum simulation of single-mode squeezing hamiltonians. %, introducing all the relevant tools of this work.
%While high-fidelity simulations can be achieved by using two-qubit interactions, the discussion of four-qubit interactions lead us naturally to Sections III and IV, where we analyze the digital quantum simulation of beam-splitters and two-mode squeezers allowing one excitation per mode. We conclude in Section V with a summary of our results.

While high-fidelity simulations can be achieved by using two-qubit interactions, the four-qubit interactions are discussed in Sections III and IV. Here,we analyze the digital quantum simulation of beam-splitters and two-mode squeezers which allows for one excitation per mode. We conclude in Section V with a summary of our results.

%In this work we study beam-splitters as an example of optical quantum system. We will simulate its dynamics, which is described by a unitary operator, and analyse the different results. It is important to mention that beam-splitters play a crucial role in boson sampling as well as in a variety of applications in continuous-variables quantum information processing and quantum computing. In fact, by Reck decomposition, we can decompose the unitary evolution governed by the hamiltonian of boson sampling into a mesh of beam-splitters and appropiate phase-shifters.

\section{Single-mode squeezing}
Squeezing is not only one of the basic processes in quantum optics but has also become a key ingredient in modern quantum technologies, from gravitational-wave astronomy \cite{gravwaves} to gaussian boson sampling \cite{supremacybs}.
Here, we discuss the digital quantum simulation of squeezing hamiltonians, starting with the single-mode case. 

As shown in \cite{losalamos, sommathesis}, it is possible to map N bosonic modes containing a maximum number of $N_p$ excitations each to $N(N_p+1)$ qubits. In the case of single-mode squeezing, we have to consider only one mode, and we can start by assuming two maximum excitations. Therefore we will need three qubits, labeled 0, 1, 2.  According to the boson-qubit mapping in \cite{losalamos,sommathesis}, we have the following Fock states  $\ket{n}$:
\begin{eqnarray}
\label{bosonmap}
|0\rangle &\leftrightarrow& |0_0 1_1 1_2 \rangle \nonumber \\
|1\rangle &\leftrightarrow& |1_0 0_1 1_2 \rangle \nonumber \\
|2\rangle &\leftrightarrow& |1_0 1_1 0_2  \rangle,
\end{eqnarray} 
where $|0_i \rangle$, $|1_i \rangle$ ($i=0, 1, 2$) are the states of qubit $i$, which are the eigenstates associated to the positive and negative eigenvalues of the Pauli operator $\sigma_z^i$, respectively.
%Particularizing 
Using the boson-qubit operator mapping for $N=1$, $N_P=2$,  we can write the bosonic creation operator as:
\begin{equation}
\label{bosonmap2}
b^{\dagger} \rightarrow 
 \sigma_-^{0}
\sigma_+^{1}+\sqrt{2} \ \sigma_-^{1}
\sigma_+^{2},
\end{equation}
where the Pauli creation and annihilation
operators are given by 
\begin{equation}\label{eq:qubitpm}
\sigma_\pm^k=\frac{1}{2} (\sigma_x^k\pm i\sigma_y^k),
\end{equation}
in terms of the Pauli matrices $\sigma_x$ and $\sigma_y$ ($k=0,1,2$). Notice that, for each qubit $\sigma_+ |0\rangle=0$, $\sigma_- |0 \rangle=|1 \rangle$, $\sigma_- |1\rangle=0$, $\sigma_+ |1 \rangle=|0 \rangle$. From Eq. (\ref{eq:qubitpm}) it is straightforward to obtain the annihilation
operator and all the combinations that subsequently appear in the bosonic Hamiltonians of interest. In particular, we want to obtain the mapping for the single-mode squeezing unitary:

%as well the annihilation operator and then all the combinations that appear in bosonic hamiltonians of interest. In particular, we want to obtain the mapping for the single-mode squeezing unitary:

\begin{equation}
U_\epsilon=e^{i\varepsilon (b^{\dagger2} + b^2)},
\end{equation}
where $\varepsilon$ is the squeezing parameter.
Using Eq. (\ref{bosonmap2}) and the properties of the Pauli operators:
\begin{eqnarray}
\sigma_{\pm}^k\sigma_{\mp}^k&=&\frac{1}{2}(1\pm\sigma_z^k)\nonumber\\
\sigma_{\pm}^k\sigma_{\pm}^k&=&0,
\end{eqnarray}
we get
\begin{eqnarray}
b^{\dagger2}&\rightarrow&\sqrt{2} \sigma_-^0\sigma_+^2\nonumber\\
b^2&\rightarrow&\sqrt{2} \sigma_+^0\sigma_-^2.
\end{eqnarray}
Then, using Eq. (\ref{eq:qubitpm}) and summing up, we get:  %some terms cancel out and we finally get:
\begin{equation}
\label{squeezingmap1}
b^{\dagger2}+b^2\rightarrow\frac{1}{\sqrt{2}} (\sigma_x^0\sigma_x^2+\sigma_y^0\sigma_y^2).
\end{equation}
With this, we have mapped a bosonic hamiltonian into a qubit one. However, we want to implement a unitary evolution in the digital simulator.
%However, what we want to implement in the digital simulator is a unitary evolution.
Notice that:
\begin{eqnarray}
\label{commute}
[\sigma_x^0\sigma_x^2,\sigma_y^0\sigma_y^2]=
[\sigma_x^0,\sigma_y^0] \sigma_x^2\sigma_y^2+\sigma_y^0\sigma_x^0[\sigma_x^2,\sigma_y^2 ]=0,
\end{eqnarray}
where in the last step we have used that $\sigma_i^k\sigma_y^k=i\epsilon_{ijk}\sigma_k^k$ ($\epsilon_{ijk}$ being the Levi-Civita tensor). Therefore, in this case the factorization of the unitary is exact:
\begin{equation}
\label{eq:unitarymap}
U_{\hat{\varepsilon}}=e^{i\hat{\varepsilon}\sigma_x^0\sigma_x^2}e^{i\hat{\varepsilon}\sigma_y^0\sigma_y^2},
\end{equation}
where $\hat{\varepsilon}=\varepsilon/\sqrt{2}$.
This unitary is not directly available in the quantum devices of IBM. Therefore, we have to perform a gate decomposition to express it in terms of the desired gate set. In general, if we define 
\begin{equation}
U_{\hat{\varepsilon}}=e^{iH\hat{\varepsilon}}
\end{equation}
and find another unitary operation $U$ such that:
\begin{equation}\label{eq:gatedec1}
H=U^{\dagger}H_0 U,
\end{equation}
where $H_0$ is  a convenient single-qubit operation, then we can write:
\begin{equation}\label{eq:gatedec2}
e^{i H\hat{\varepsilon}}=U^{\dagger}e^{i H_0 \hat{\varepsilon}} U.
\end{equation}

For instance, for the XX part we can use:
\begin{eqnarray}
e^{-i\frac{\pi}{4}\sigma_x^{0}}(\sigma_z^{0})e^{i\frac{\pi}{4}\sigma_x^{0}}&=&-\sigma_y^{0}\\\nonumber
e^{-i\frac{\pi}{4}\sigma_z^{0}\sigma_x^{2}}(-\sigma_y^{0})e^{i\frac{\pi}{4}\sigma_z^{0}\sigma_x^{2}}&=&\sigma_x^{0}\sigma_x^{2}.
\end{eqnarray}
Then:
\begin{equation}
e^{i\hat{\varepsilon}\sigma_x^0\sigma_x^2}=U_{xx}^{\dagger}e^{i\hat{\varepsilon}\sigma_z^0}U_{xx},
\end{equation}
where
\begin{equation}
U_{xx}=e^{i\frac{\pi}{4}\sigma_x^{0}}e^{i\frac{\pi}{4}\sigma_z^{0}\sigma_x^{2}}.
\end{equation}
Similarly, we can write:
\begin{equation}
e^{i\hat{\varepsilon}\sigma_y^0\sigma_y^2}=U_{yy}^{\dagger}e^{i\hat{\varepsilon}\sigma_z^0}U_{yy},
\end{equation}
where
\begin{equation}
U_{yy}=U_{xx}e^{i\frac{\pi}{4}\sigma_z^0}e^{i\frac{\pi}{4}\sigma_z^2}.
\end{equation}
Finally, the two-qubit gates can be translated into CNOT gates by using:
\begin{equation}
e^{i\frac{\pi}{4}\sigma_z^{0}\sigma_x^{2}}= e^{i\frac{\pi}{4}\sigma_x^2}e^{i\frac{\pi}{4}\sigma_z^{0}}e^{-i\frac{\pi}{4}} CNOT^{0-2} \label{eq:cnot},
\end{equation}
where the CNOT gate between a pair of qubits $i$, $j$ is defined as:
\begin{equation}
CNOT^{i-j}= \begin{pmatrix}1&0 &0 &0\\0&0&0 &1 \\ 0&0 &1 &0 \\ 0&1 &0 &0\end{pmatrix}.
\end{equation}

We also define the single-qubit single-parameter rotation
\begin{equation}
U_1(\lambda)= \begin{pmatrix}1&0\\0&e^{i\lambda}\end{pmatrix},
\end{equation}
and the single-qubit two-parameter rotation:
\begin{equation}
U_2(\phi,\lambda)= \frac{1}{\sqrt{2}}\begin{pmatrix}1&-e^{i\lambda}\\e^{i\phi}&e^{i\lambda+i\phi}\end{pmatrix},
\end{equation}
using the same conventions as IBM.

We note that
\begin{eqnarray}
e^{i\frac{\pi}{4}\sigma_x^0}e^{i\frac{\pi}{4}\sigma_z^0}&=&e^{i\frac{\pi}{4}}U_2^0(\frac{\pi}{2},-\pi)\nonumber\\
e^{-i\frac{\pi}{4}\sigma_z^0}e^{-i\frac{\pi}{4}\sigma_x^0}&=&e^{-i\frac{\pi}{4}}U_2^0(0,\frac{\pi}{2})\nonumber\\
e^{i\sigma_z^k\theta}&=&e^{i\theta}U_1^k(-2\theta).
\end{eqnarray}
Putting everything together, phases and some operations on qubit 2 cancel out and we get a final expression with 10 single-qubit gates and four CNOT gates:
\begin{figure*}
    \includegraphics[width=\textwidth]{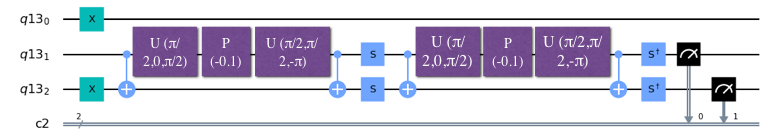}
    \caption{Quantum circuit run in Santiago for the single-mode squeezing unitary Eq. (\ref{eq:unitarysim}) and $\hat{\varepsilon}=0.05$.
    $q13_0$, $q13_1$ and $q13_2$ are respectively qubits 1, 0 and 2 in the text.}
    \label{fig:santiagoaa} 
\end{figure*}
\begin{figure*}
    \includegraphics[width=0.8\textwidth]{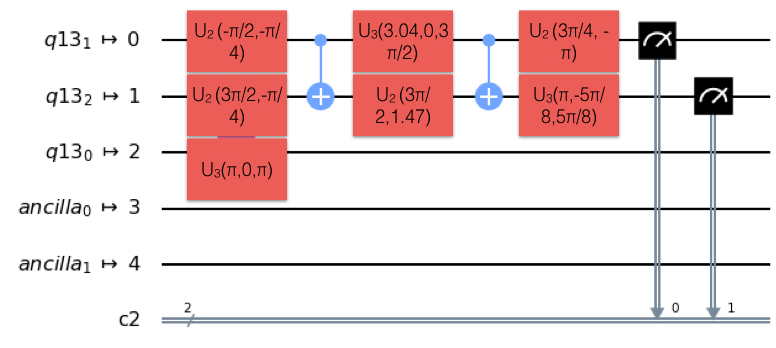}
    \caption{Final transpiled circuit run in Santiago for the single-mode squeezing unitary Eq. (\ref{eq:unitarysim}) and $\hat{\varepsilon}=0.05$.}
    \label{fig:CASABLANCAAaa} 
\end{figure*}
\begin{eqnarray} \label{eq:unitarysim}
U_{\hat{\varepsilon}}&=&CNOT^{0-2}U_2^0\left(0,\frac{\pi}{2}\right)U_1^0\left(-2\hat{\varepsilon}\right)U_2^0\left(\frac{\pi}{2},-\pi\right)\nonumber\\&&CNOT^{0-2}U_1^2\left(\frac{\pi}{2}\right)U_1^0\left(\frac{\pi}{2}\right)\nonumber\\&&CNOT^{0-2}U_2^0\left(0,\frac{\pi}{2}\right)U_1^0\left(-2\hat{\varepsilon}\right)U_2^0\left(\frac{\pi}{2},-\pi\right)\nonumber\\&&CNOT^{0-2}U_1^2\left(-\frac{\pi}{2}\right)U_1^0\left(-\frac{\pi}{2}\right).
\end{eqnarray}

The $U_1(\pi/2)$ gate will be denoted from now on as a $S$ gate.
Of course, the product of several single-qubit operations on the same qubit are effectively grouped into a single operation when launched into an IBM quantum device. In Fig. (\ref{fig:santiagoaa}), we see the circuit corresponding to the unitary Eq. (\ref{eq:unitarysim}) for $\hat{\varepsilon}=0.05$ acting on an initial ground state as launched by us in IBM Santiago and in Fig. (\ref{fig:CASABLANCAAaa}) the final ``transpiled'' version. Fortunately, the number of CNOT gates is reduced from 4 to 2. All the circuits were transpiled using qiskit optimization level 3 \cite{qiskit}. Note that the two X-gates in Fig. \ref{fig:santiagoaa} are required for the preparation of the initial ground state and that in Fig. \ref{fig:CASABLANCAAaa} the single-qubit three-parameter rotation is:
\begin{equation}
U_3(\theta,\phi,\lambda)= \frac{1}{\sqrt{2}}\begin{pmatrix}\cos\left(\frac{\theta}{2}\right)&-e^{i\lambda}\sin\left(\frac{\theta}{2}\right)\\e^{i\phi}\sin\left(\frac{\theta}{2}\right)&e^{i\lambda+i\phi}\cos\left(\frac{\theta}{2}\right)\end{pmatrix}.
\end{equation}
 %the corresponding codes are attached as additional material \cite{supp}. 

Notice that the restriction to a maximum of two photons corresponds to a restriction to perturbative values of $\varepsilon$. In second-order perturbation theory, the state that we intend to simulate would have the form $|\psi\rangle=(1-\varepsilon^2)|0\rangle-i\,\sqrt{2}\varepsilon|2\rangle$.
Our aim is to compute the fidelity 
\begin{eqnarray}
    F(\left\lvert \psi\right\rangle,\rho)=\left\langle\psi\right\rvert\rho\left\lvert\psi\right\rangle.
\end{eqnarray}
where $\psi$ is the aforementioned perturbative state and $\rho$ is the state actually obtained in the experiment. 
We would like to perform a full state tomography of $\rho$. An arbitrary state of $n$ qubits can be expanded as \cite{nielsenchuang} 
\begin{equation}\label{tomography}
    \rho=\sum_{\Vec{v}}\frac{\text{tr}(\sigma_{v_1}\otimes\cdots\otimes\sigma_{v_n}\rho)\;\sigma_{v_1}\otimes\cdots\otimes\sigma_{v_n}}{2^n}\;,
\end{equation}
where the sum is over vectors $\Vec{v}=(v_1,\dots,v_n)$ with entries $v_i=0,1,2,3$. Considering Eq. (\ref{tomography}) for three qubits and the expression of $\psi$, we arrive at the following expression for the fidelity in terms of $\varepsilon$. 
%Particularizing for three qubits and considering the expression of $\psi$, we arrive at the following expression for the fidelity in terms of $\varepsilon$. 

\begin{eqnarray}\label{eq:monster}
F(\left\lvert \psi\right\rangle,\rho&&)=\frac{1}{8}\Big(1+\operatorname{tr}(\sigma_{z}\otimes\sigma_{z}\otimes\sigma_{z}\rho)+2\sqrt{2}\varepsilon\nonumber\\&&\big[\operatorname{tr}(\sigma_{x}\otimes\sigma_{z}\otimes\sigma_{y}\rho)+\operatorname{tr}(\sigma_{y}\otimes\operatorname{Id}\otimes\sigma_{x}\rho)\nonumber\\&&-\operatorname{tr}(\sigma_{x}\otimes\operatorname{Id}\otimes\sigma_{y}\rho)-\operatorname{tr}(\sigma_{y}\otimes\sigma_{z}\otimes\sigma_{x}\rho)\big]+\nonumber\\&&(1-4\varepsilon^2)\big[\operatorname{tr}(\sigma_{z}\otimes\operatorname{Id}\otimes\operatorname{Id}\rho)+\operatorname{tr}(\operatorname{Id}\otimes\sigma_{z}\otimes\sigma_{z}\rho)\nonumber\\&&-\operatorname{tr}(\operatorname{Id}\otimes\operatorname{Id}\otimes\sigma_{z}\rho)-\operatorname{tr}(\sigma_{z}\otimes\sigma_{z}\otimes\operatorname{Id}\rho)\big]\nonumber\\&&-\big[\operatorname{tr}(\operatorname{Id}\otimes\sigma_{z}\otimes\operatorname{Id}\rho)+\operatorname{tr}(\sigma_{z}\otimes\operatorname{Id}\otimes\sigma_{z}\rho)\big]\Big).
\end{eqnarray}

Full state tomography is not straightforward in IBM quantum devices since only measurements in the Z basis are allowed. However, we can simulate the measurements in the X and Y basis by adding a Hamadard gate for the former and $S^{\dagger}$ and Hadamard for the latter, prior to the standard measurement. We recall that the Hadamard gate is defined as:
\begin{equation}
H= \frac{1}{\sqrt{2}}\begin{pmatrix}1&1\\1&-1\end{pmatrix}.
\end{equation}

Therefore, we can repeat the experiment several times, each time performing the measurements in the corresponding basis, in order to obtain all the terms needed to compute the fidelity. Alternatively, we can launch a single experiment, and --instead of performing state tomography-- make use of the fact that the leading order term of the fidelity would be 
\begin{equation}\label{eq:fidapprox}
F=P_0, 
\end{equation}
where $P_0$ is the probability of the ground state, which can be easily retrieved from the experiment with qiskit, thus obtaining a second-order approximation error. Moreover, since not all the Hilbert space of the three qubits is associated with physical states in the simulation, we can also use post-selection, by simply neglecting all the probability counts of the states that are not related with the simulated Fock states in Eq. (\ref{bosonmap}). However, in this case the post-selection probabilities are very close to 1, so there is almost no effect in the results.

In all the cases, we also make use of error mitigation techniques \cite{mitigation}, which in the case of the IBM devices are only available for the measurements. 

\begin{figure}
\includegraphics[width=0.5\textwidth]{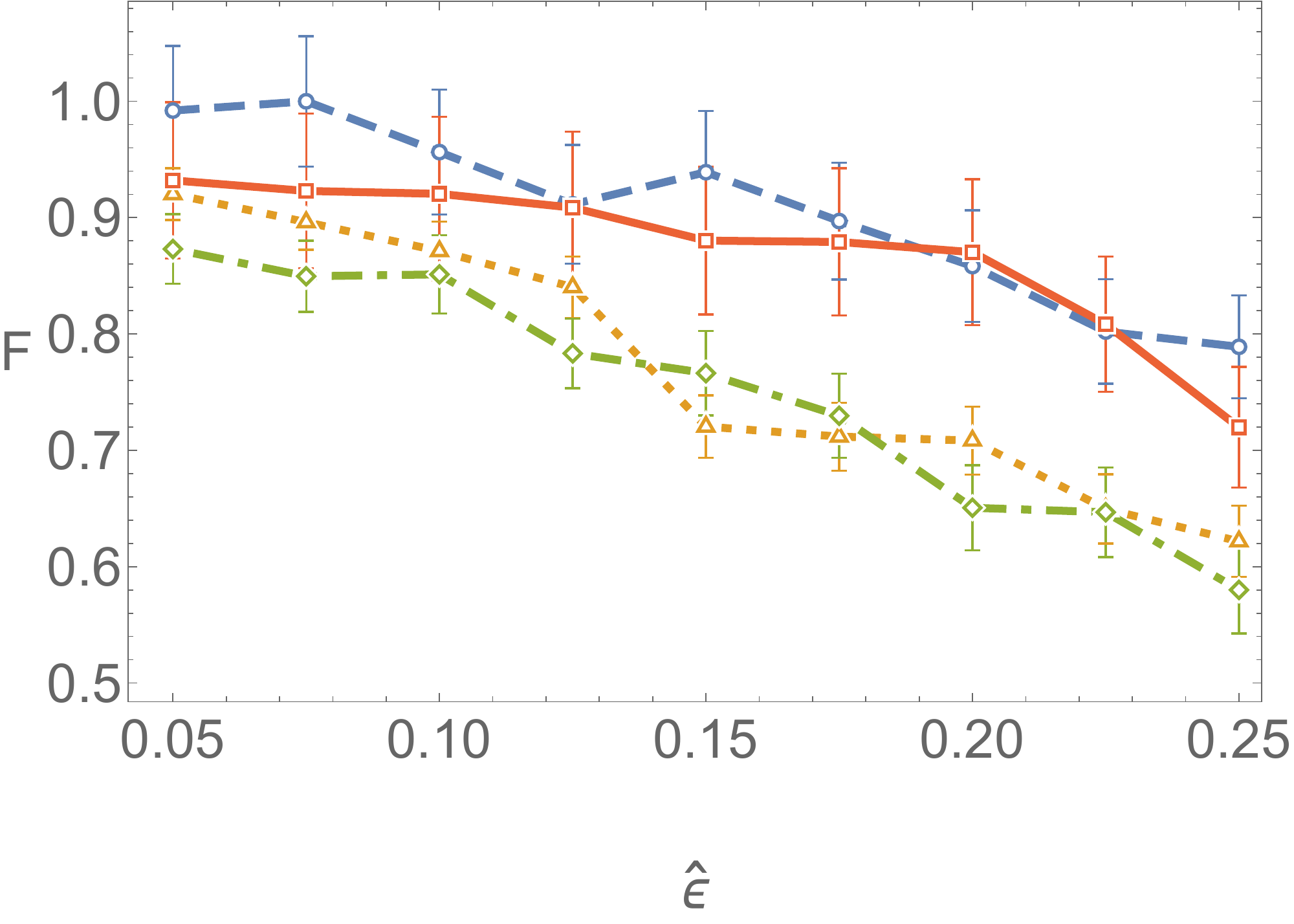}
 \caption{Fidelity $F$ results for the digital quantum simulation of single-mode squeezing with squeezing parameter $\hat{\varepsilon}$ allowing a maximum of two excitations for: Santiago (May/04/2021), using the approximation in Eq.(\ref{eq:fidapprox}) (blue, dashed, circles) and with full tomography (May/19-20/2021) Eq.(\ref{eq:monster}) (orange, dotted, triangles) and Casablanca (May/05/2021), Eq. (\ref{eq:fidapprox}) (red, solid, squares) and (May/19/2021) Eq. (\ref{eq:monster}) (green, dash-dotted, diamonds). }
    \label{fig:1msq2exc} 
\end{figure}

In Fig. \ref{fig:1msq2exc}, we show the results of the simulation for different parameters and devices.
Throughout this work, we will present results in two IBM quantum devices, namely Casablanca and Santiago. They both display a high quantum volume $\operatorname{QV}=32$ \cite{quantumvolume} and, in the case of Casablanca, the required connectivity, that is, the availability of CNOT gates, for all the simulations in this work. While the connectivity of Santiago is also enough by now, that will not be the case in the next sections. The non-availability of a CNOT gate implies the introduction of additional SWAP operations in order to move the CNOT gate to another pair of qubits where the CNOT can be implemented, which would have an important impact in the fidelity. Typical error rates in Santiago range from $6.4\times10^{-3}$ to $2.36\times 10^{-2}$ for readout (note however that we have used error mitigation for the readout), from $1.7\times 10^{-4}$ to $2.6\times 10^{-4}$ for single-qubit gates and from $5.5\times 10^{-3}$ to $6.5\times 10^{-3}$ for CNOT gates. While in Casablanca typical errors go from $9.4\times 10^{-3}$ to $6.2\times 10^{-2}$ for readout, from $2.5\times 10^{-4}$ to $6.9\times 10^{-3}$ for single-qubit gates and from 8$\times 10^{-3}$ to $6.2\times 10^{-2}$ for the CNOT gates. Error bars can be assigned by considering a typical average readout assignment error -$1.4\times10^{-2}$ in Santiago and $1.8\times10^{-2}$ in Casablanca-  and standard error propagation techniques. Note however that all these parameters change on a daily basis. Therefore we will include the date when the results were obtained throughout the text. We have also used other quantum devices (results not included) and checked that there is significant dependence of the fidelity with QV and connectivity.  It is important to mention that the qiskit version used in all the simulations was '0.21.0'.

With only two CNOT gates and readouts and a few single-qubit gates, if the error is mainly due to the error gate -as should be for small $\hat{\varepsilon}$, where our approximation should be very close to the exact dynamics- with the numbers above fidelities above 95 \% can be expected in a single experiment, as can be seen in Fig. \ref{fig:1msq2exc}. Of course, the repetitions of the experiment in order to achieve full tomography have a negative impact in the fidelity. As expected, we also see in Fig. \ref{fig:1msq2exc} that the fidelity decreases for larger $\hat{\varepsilon}$, since the perturbative approximation starts to fail.
If we want to simulate higher values of the squeezing parameter we need to allow a higher number of photons. 
 Taking four photons, we will now need 5 qubits, and the new definition of the creation operator would be:
\begin{equation}
\label{eq:bosonmap4phot} 
 b^{\dagger} \rightarrow
 \sigma_-^{0}
\sigma_+^{1}+\sqrt{2} \ \sigma_-^{1}
\sigma_+^{2}+\sqrt{3} \ \sigma_-^{2}
\sigma_+^{3}+2 \ \sigma_-^{3}
\sigma_+^{4}.
\end{equation}
With this definition $(b^{\dagger})^2+b^2$ is mapped to a qubit operator containing both two-qubit interactions $S_2$ and four-qubit ones $S_4$. Let us analyse first the two-qubit part $S_2$, which is:
\begin{eqnarray}\label{eq:s2}
S_2&=&\frac{1}{\sqrt{2}} (\sigma_x^0\sigma_x^2+\sigma_y^0\sigma_y^2)+\sqrt{3}(\sigma_x^2\sigma_x^4+\sigma_y^2\sigma_y^4)+\nonumber\\&&\sqrt{\frac{3}{2}}(\sigma_x^1\sigma_x^3+\sigma_y^1\sigma_y^3),
\end{eqnarray}
where the first term is exactly the same as in the two-photon scenario, and the other two only differ in the qubit pair onto which they are applied. However, a crucial feature is that now the second term in $S_2$ does not commute with the first one. 
Then, Trotter decomposition techniques can be useful \cite{lloyd}.

\begin{figure*}
\includegraphics[width=\textwidth]{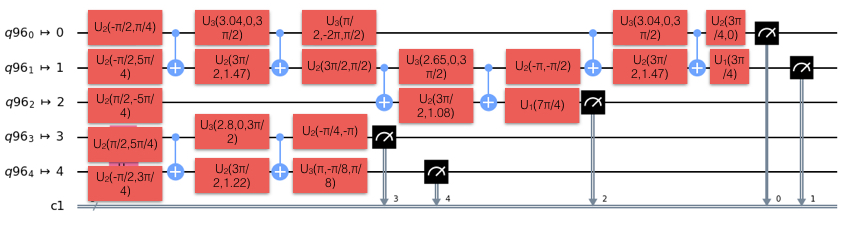} 
 \caption{Final transpiled circuit run in ibmq$\_$santiago for the single-mode squeezing unitary allowing 4 excitations and $\hat{\varepsilon}=0.1$. We consider two-qubit interactions only, as in Eq. (\ref{eq:s2}). $q96_0$, $q96_1$, $q96_2$, $q96_3$ and $q96_4$ are respectively qubits 0, 3, 1, 4 and 2 in the text.}
    \label{fig:CASABLANCAAadda} 
\end{figure*}

Indeed, we use two alternate approaches. On one hand, we use the Trotter formula \cite{lloyd,nielsenchuang}
\begin{eqnarray}\label{eq:stepsn}
   e^{i(A+B)\hat{\varepsilon}}=\lim_{n\to\infty}(e^{iA\hat{\varepsilon}/n}e^{iB\hat{\varepsilon}/n})^n,
\end{eqnarray}
where $n$ is the number of Trotter steps, $A$ and $B$ are two non-commuting parts of Eq. (\ref{eq:s2})  and we are treating $\hat{\varepsilon}$ as the time variable.
Alternatively, we also make use  of the following higher-order approximation \cite{nielsenchuang}:
\begin{equation}\label{eq:order3}
    e^{i(A+B)\hat{\varepsilon}}=e^{iA\hat{\varepsilon}/2}e^{iB\hat{\varepsilon} }e^{iA\hat{\varepsilon}/2}+\mathcal{O}(\hat{\varepsilon}^3).
\end{equation}
We see in Fig. (\ref{fig:CASABLANCAAadda}) the corresponding circuit for the latter.

In this case, it is not feasible to calculate the fidelity using the full state tomography as we will need to compute up to 1024 terms. So we just calculate the fidelity up to second-order corrections, by using Eq. (\ref{eq:fidapprox}). 
Moreover, we post-select the states $\left\lvert 01111\right\rangle, \left\lvert 10111\right\rangle, \left\lvert 11011\right\rangle,  \left\lvert 11101\right\rangle$ and $ \left\lvert 11110\right\rangle$.

\begin{figure}
    \centering
    \includegraphics[width=0.5\textwidth]{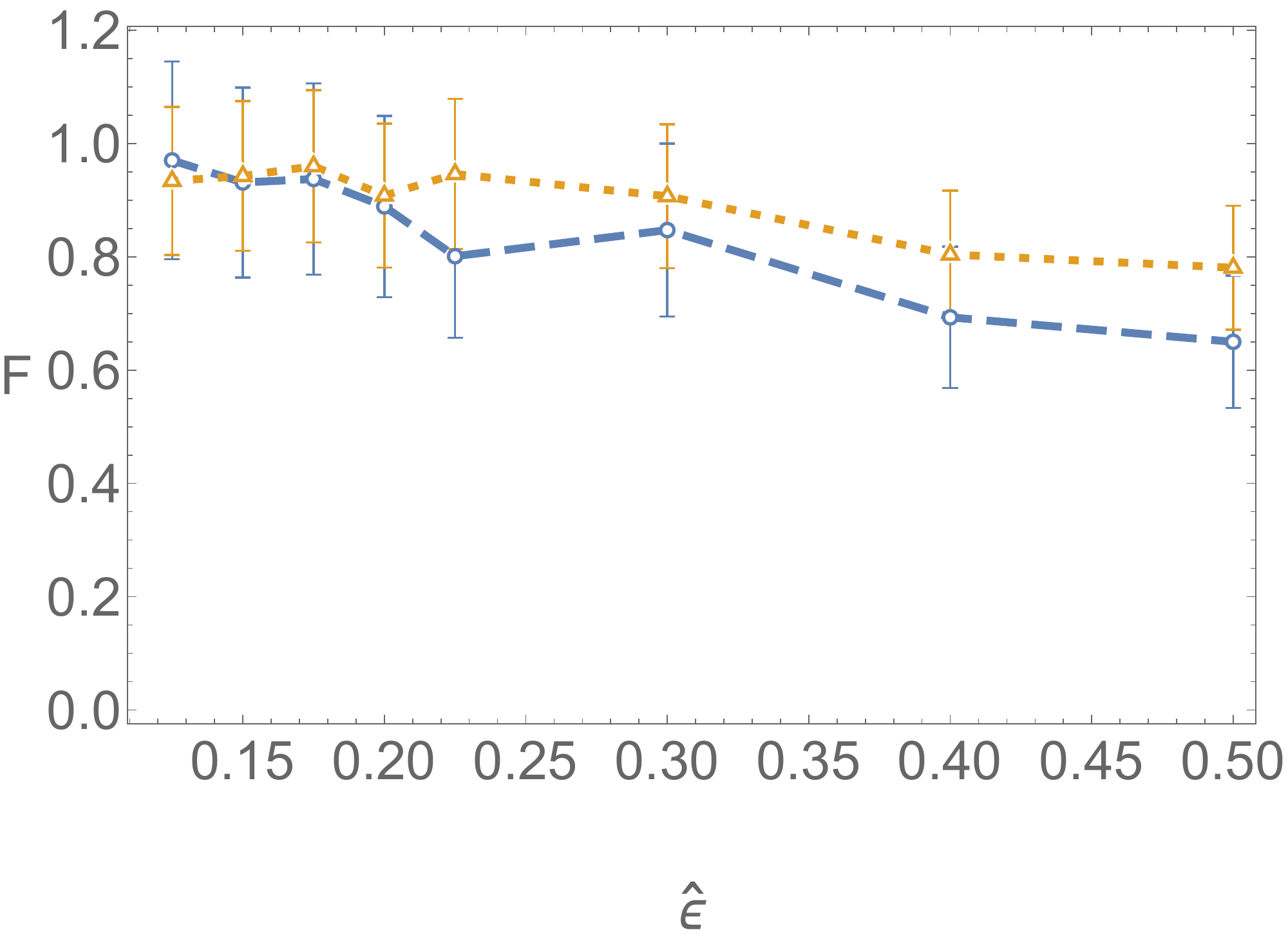}
    \caption{Fidelity $F$ results for the digital quantum simulation of single-mode squeezing with squeezing parameter $\hat{\varepsilon}$ allowing a maximum of four excitations by using Eq. (\ref{eq:order3}): Casablanca (May/05-06/2021) (blue, dashed, circles) and Santiago (May/20-21/2021) (orange, dotted, triangles).}
    \label{fig:1msq4exc} 
\end{figure}

\begin{figure}
    \centering
    \includegraphics[width=0.5\textwidth]{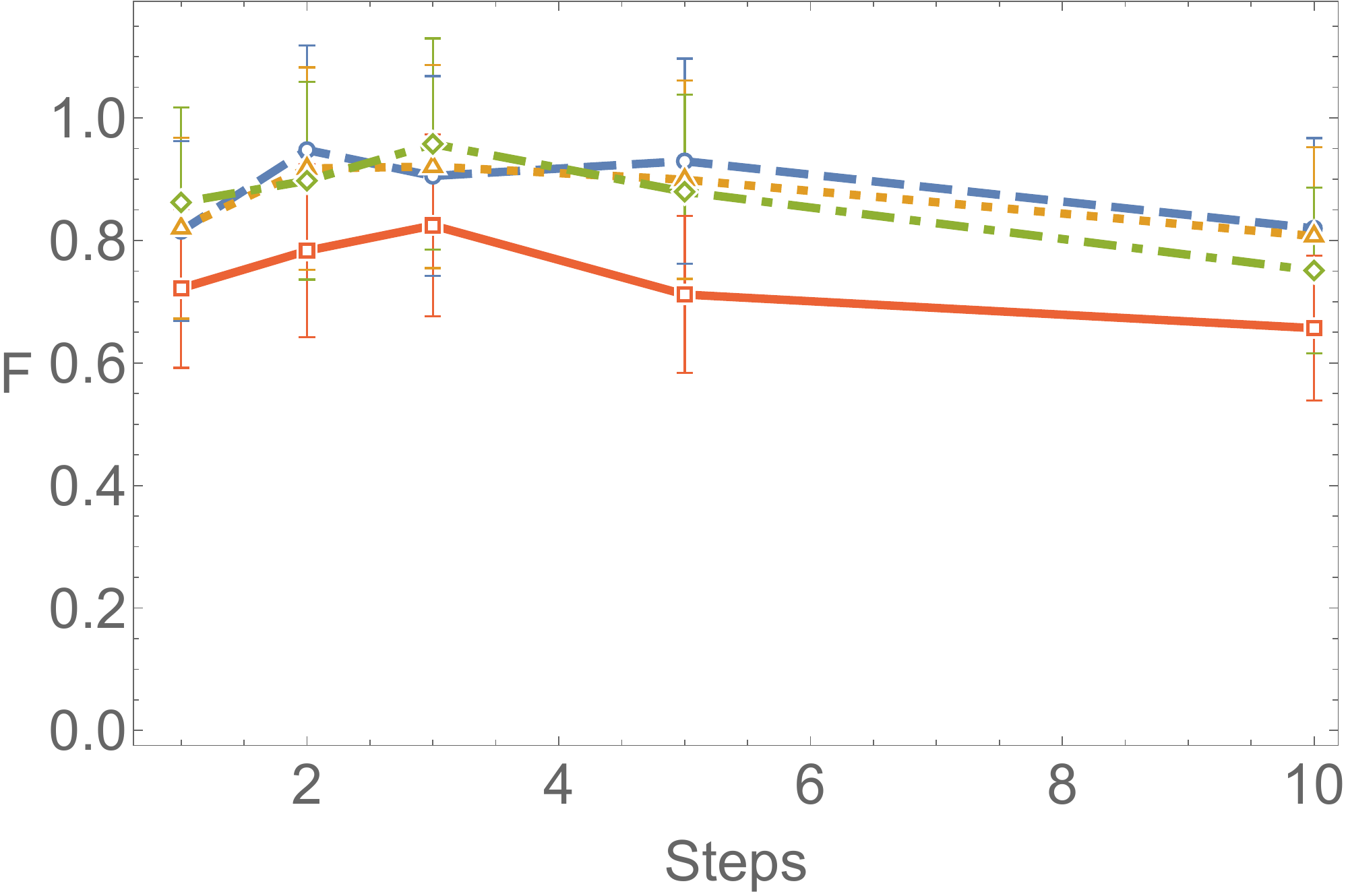}
    \caption{Fidelity $F$ vs. number of Trotter steps for the digital quantum simulation of single-mode squeezing allowing a maximum of four excitations in  Casablanca  (May/05-06/2021) for $\hat{\varepsilon}=0.125$(blue, dashed, circles) $\hat{\varepsilon}=0.175$ (orange, dotted, triangles) $\hat{\varepsilon}=0.2$ (green, dash-dotted, diamonds) and $\hat{\varepsilon}=0.3$  (red, solid, squares) .}
    \label{fig:1msq4excsteps} 
\end{figure}

\begin{figure}
    \centering
    \includegraphics[width=0.5\textwidth]{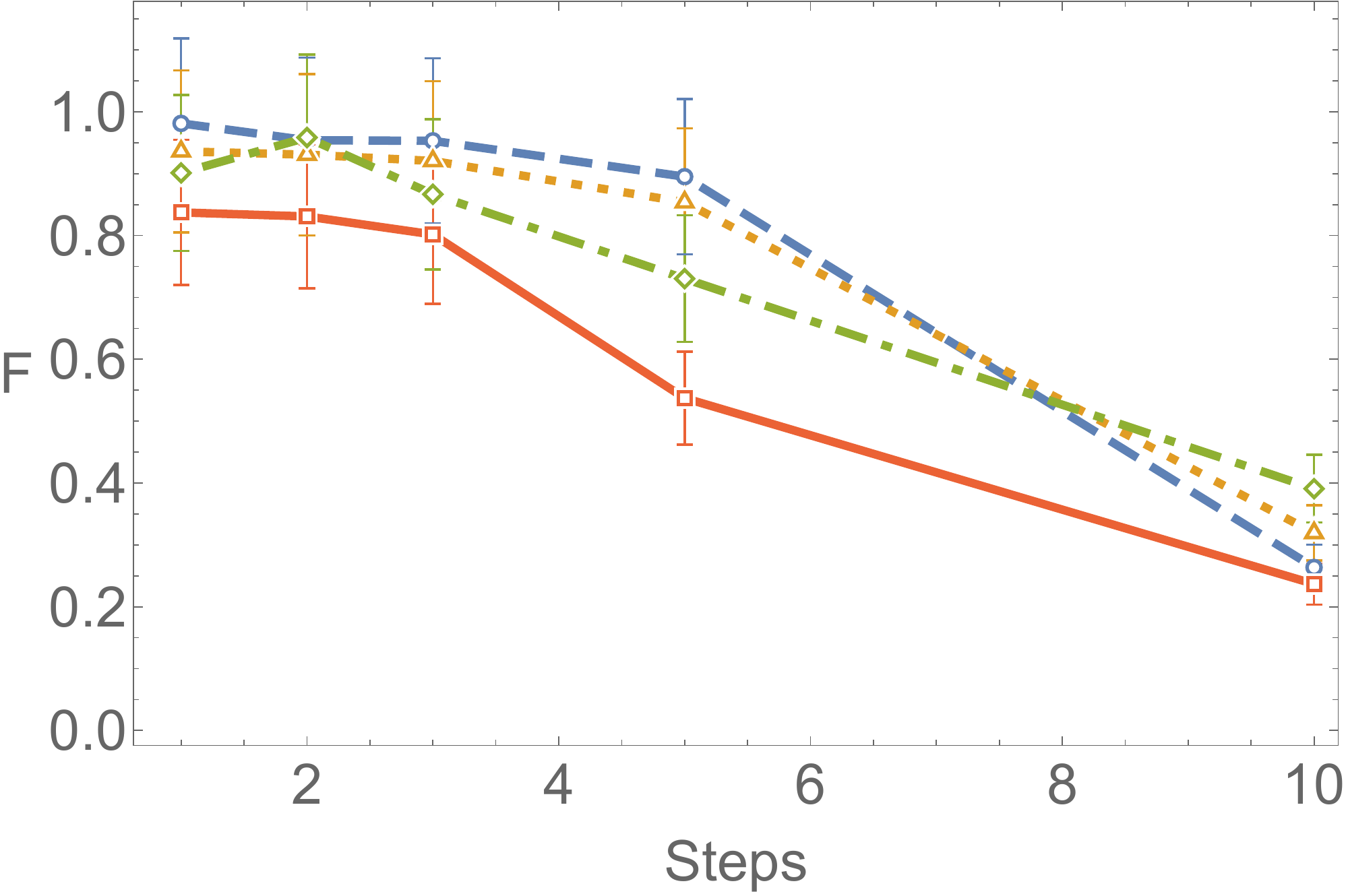}
    \caption{Fidelity $F$ vs. number of Trotter steps for the digital quantum simulation of single-mode squeezing allowing a maximum of four excitations in Santiago (May/20-21/2021)   for $\hat{\varepsilon}=0.125$(blue, dashed, circles) $\hat{\varepsilon}=0.175$ (orange, dotted, triangles) $\hat{\varepsilon}=0.2$ (green, dash-dotted, diamonds) and $\hat{\varepsilon}=0.3$  (red, solid, squares) .}
    \label{fig:1msq4excstepssanti} 
\end{figure}

Putting everything together, we show the fidelity results in Figs. \ref{fig:1msq4exc}, \ref{fig:1msq4excsteps} and \ref{fig:1msq4excstepssanti}. In Fig \ref{fig:1msq4exc} we see that the use of Eq. (\ref{eq:order3}) allows us to extend the large-fidelity region to larger values of $\hat{\varepsilon}$. Also in Figs. \ref{fig:1msq4excsteps} and \ref{fig:1msq4excstepssanti} we see that the fidelity increases for a low number of Trotter steps, specially in Casablanca and for moderate values of the squeezing. However, for large values of the squeezing parameter and a large number of Trotter steps the error due to barriers and gates suppress the benefits of trotterization. 

So far, we have not considered the four-qubit term $S_4$,which is:
\begin{eqnarray}
\label{eq:s4term}
S_4&=& 2\sqrt{3}(\sigma_-^{0}\sigma_+^{1}\sigma_-^{2}\sigma_+^{3}+\sigma_+^{0}\sigma_-^{1}\sigma_+^{2}\sigma_-^{3})\nonumber\\&+&4\sqrt{2}(\sigma_-^{1}\sigma_+^{2}\sigma_-^{3}\sigma_+^{4}+\sigma_+^{1}\sigma_-^{2}\sigma_+^{3}\sigma_-^{4})\nonumber\\&+& 4(\sigma_-^{0}\sigma_+^{1}\sigma_-^{3}\sigma_+^{4}+\sigma_+^{0}\sigma_-^{1}\sigma_+^{3}\sigma_-^{4}), 
\end{eqnarray}
where each of the three contributions gives rise to
\begin{eqnarray}
\label{eq:s4sigmas}
& &\sigma_-^{i}\sigma_+^{j}\sigma_-^{k}\sigma_+^{l}+\sigma_+^{i}\sigma_-^{j}\sigma_+^{k}\sigma_-^{l}=\nonumber\\& &\frac{1}{8}(\sigma_x^{i}\sigma_x^{j}\sigma_x^{k}\sigma_x^{l}+\sigma_x^{i}\sigma_y^{j}\sigma_y^{k}\sigma_x^{l}-\sigma_x^{i}\sigma_y^{j}\sigma_x^{k}\sigma_y^{l}+\sigma_x^{i}\sigma_x^{j}\sigma_y^{k}\sigma_y^{l}+\nonumber \\& &\sigma_y^{i}\sigma_y^{j}\sigma_x^{k}\sigma_x^{l}-\sigma_y^{i}\sigma_x^{j}\sigma_y^{k}\sigma_x^{l}+\sigma_y^{i}\sigma_x^{j}\sigma_x^{k}\sigma_y^{l}+\sigma_y^{i}\sigma_y^{j}\sigma_y^{k}\sigma_y^{l}).
\end{eqnarray}
However, we have seen that we are able to achieve high-fidelity simulations without including them --especially for low squeezing-- while for high squeezing we should consider a larger number of allowed excitations. Therefore, instead to include these terms in the single-mode squeezing simulations, we analyse them separately. The four-qubit operators in Eq. (\ref{eq:s4sigmas}) are exactly the same as a full two-mode squeezing hamiltonian $a^{\dagger}b^{\dagger}+a b$ for two modes $a$, $b$ with a maximum of one excitation per mode allowed. Thus, later on we will discuss an interesting application of two-mode squeezing with a maximum of one photon per mode, such as the dynamical Casimir effect \cite{dce}, where starting with an initial vacuum state we have a certain probability of generating a photon pair by means of the modulation of boundary conditions. However, we will start with the related quantum digital simulation of a beam-splitter hamiltonian.

\section{Beam-splitters}

The recipe for the digitalization of four-qubit interactions similar to those in Eq. (\ref{eq:s4sigmas}) was discussed in \cite{sabindqs} in the context of beam-splitting hamiltonians. A beam-splitter hamiltonian for two modes with a maximum of one photon per mode gives rise to a similar series of four-qubit interaction terms as Eq.(\ref{eq:s4sigmas}), with the only difference of some of the signs.

Labelling the two modes of interest as $+$ and $-$, each of the modes is represented by two qubits and, since $N_P=1$ the bosonic creation operator consists just of the first term of Eq. (\ref{bosonmap2}). Thus, the beam-splitter unitary could be written as:
\begin{equation}
    U_{+-}=e^{i\varepsilon_{+-}(b_+^{\dag}a_-+ h.c.)}\;,
\end{equation}
Note that such unitary evolution can be exactly solved. In our particular case, the evolution of the initial state with a photon in one mode is
\begin{eqnarray}\label{final}
    \left\lvert \psi\right\rangle&=&U_{+-}\left\lvert 1\right\rangle_+\left\lvert 0\right\rangle_-=\cos{\varepsilon_{+-}}\left\lvert 1\right\rangle_+\left\lvert 0\right\rangle_-+\nonumber\\&&i\sin{\varepsilon_{+-}}\left\lvert 0\right\rangle_+\left\lvert 1\right\rangle_-.
\end{eqnarray}
This will enable us to compute the fidelity in an exact fashion later on. For the moment, we will digitalize the bosonic operator
\begin{eqnarray}\label{eq:mappedbs}
& &b_+^{\dag}a_-+ b_+a_-^{\dag} \rightarrow\frac{1}{8}(\sigma_x^{0}\sigma_x^{1}\sigma_x^{2}\sigma_x^{3}-\sigma_x^{0}\sigma_y^{1}\sigma_y^{2}\sigma_x^{3}\nonumber\\ & &+\sigma_x^{0}\sigma_y^{1}\sigma_x^{2}\sigma_y^{3}+\sigma_x^{0}\sigma_x^{1}\sigma_y^{2}\sigma_y^{3}+\sigma_y^{0}\sigma_y^{1}\sigma_x^{2}\sigma_x^{3}\nonumber \\& &+\sigma_y^{0}\sigma_x^{1}\sigma_y^{2}\sigma_x^{3}-\sigma_y^{0}\sigma_x^{1}\sigma_x^{2}\sigma_y^{3}+\sigma_y^{0}\sigma_y^{1}\sigma_y^{2}\sigma_y^{3}).
\end{eqnarray}
This operator can be expressed as a product of simpler unitaries:
\begin{equation}
    U_{+-}=\prod_{i=1}^{8} U_{+-}^{i}\;,
\end{equation}
where each $U_{+-}^{i}$ is easily decomposed into basic gates. For instance:
\begin{equation}\label{1}
    U_{+-}^{1}=U^{\dag} e^{-i\frac{\epsilon_{+-}}{8}\sigma_z^{0}} U,
\end{equation}
where:
\begin{equation}\label{unitary}
    U=e^{i\frac{\pi}{4}\sigma_x^{0}}e^{i\frac{\pi}{4}\sigma_z^{0}\sigma_x^{1}}  e^{i\frac{\pi}{4}\sigma_z^{0}\sigma_x^{2}}e^{i\frac{\pi}{4}\sigma_z^{0}\sigma_x^{3}}\;.
\end{equation}
Note that similar decompositions can be obtained for the rest of $U_{+-}^{i}$'s by adding at the end of the string the number of $e^{i\pi/4\sigma_z^{j}}$, necessary to rotate some of the $\sigma_x$ to $\sigma_y$.
Putting all the above together, a single two-mode beam-splitter with one photon per mode can be simulated in a four-qubit quantum simulator. The four possible quantum states are mapped  in the following way:
\begin{align*}
    \left\lvert 0\right\rangle_+\left\lvert 0\right\rangle_-&=\left\lvert 0101\right\rangle\\
    \left\lvert 0\right\rangle_+\left\lvert 1\right\rangle_-&=\left\lvert 0110\right\rangle\\
    \left\lvert 1\right\rangle_+\left\lvert 0\right\rangle_-&=\left\lvert 1001\right\rangle\\
    \left\lvert 1\right\rangle_+\left\lvert 1\right\rangle_-&=\left\lvert 1010\right\rangle.
\end{align*}
% These states can be obtained by spin-flipping a pair of qubits out of the four-qubit ground state. Choosing the initial state $\left\lvert 1\right\rangle_+\left\lvert 0\right\rangle_-$ and applying the unitary dynamics we get:

\begin{figure*}
    \centering
    \includegraphics[width=\textwidth]{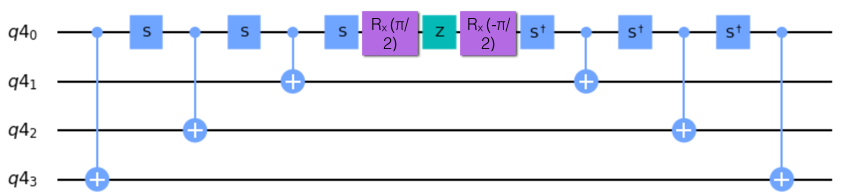}
    \caption{Sequence of gates that implement the beam-splitting interaction term $U_{+-}^{1}$ for $\tfrac{\varepsilon_{+-}}{8}=\tfrac{\pi}{2}$. The order of the qubits is the same as in the text.}
    \label{fig:XX} 
\end{figure*}

Finally, to have everything ready for the simulation we need to rewrite Eq. (\ref{unitary}) in terms of the gates available in IBM architecture, by using similar techniques as in the previous section. The circuit in Fig. (\ref{fig:XX}) simulates the first block $U_{+-}^{1}$ for $\tfrac{\varepsilon_{+-}}{8}=\tfrac{\pi}{2}$, except for an irrelevant global phase. Notice that in this case a higher level of connectivity is required: one qubit needs to be connected via CNOT gates with three different qubits.

Note that, unlike in the previous section, we can now consider that each mode is a qubit with two possible states $\left\lvert 0\right\rangle$ and $\left\lvert 1\right\rangle$, although each mode is simulated by two physical qubits. Using Eq. (\ref{tomography}) for a two-qubit Hilbert space and with some algebra, we have:
\begin{eqnarray}\label{fidel}
  F(\left\lvert \psi\right\rangle,\rho)&=&\nonumber\\ \frac{1}{4}&\sum_{i,j}&\operatorname{tr}(\sigma_{i,+}\otimes\sigma_{j,-}\rho)\left\langle\psi\right\rvert\sigma_{i,+}\otimes\sigma_{j,-}\left\lvert\psi\right\rangle
\end{eqnarray}
where $i,j=0,1,2,3$. The identity and Pauli matrices $\sigma_{i,+}$, $\sigma_{j,-}$ act over the bosonic states, i.e. $\sigma_{1,+}\left\lvert 0\right\rangle_+=\left\lvert 1\right\rangle_+$, and not over individual qubits. To relax notation, we will just write $\sigma_{i}$ , $\sigma_{j}$.

In all the simulations launched, our initial state is $\left\lvert 1\right\rangle_+\left\lvert 0\right\rangle_-$. So doing straightforward calculations, we obtain that the only non-zero  terms from the prior expression are:
\begin{gather*}
    \left\langle\psi\right\rvert\sigma_z\otimes\;\operatorname{Id}\;\left\lvert\psi\right\rangle=1-2\cos^2{\varepsilon_{+-}}\\
    \left\langle\psi\right\rvert\;\operatorname{Id}\;\otimes\sigma_z\left\lvert\psi\right\rangle=-(1-2\cos^2{\varepsilon_{+-}})\\
    \left\langle\psi\right\rvert\sigma_x\otimes\sigma_y\left\lvert\psi\right\rangle=\sin{2\varepsilon_{+-}}\\
    \left\langle\psi\right\rvert\sigma_y\otimes\sigma_x\left\lvert\psi\right\rangle=-\sin{2\varepsilon_{+-}}\\
    \left\langle\psi\right\rvert\sigma_z\otimes\sigma_z\left\lvert\psi\right\rangle=-1\;.
\end{gather*}

Keeping in mind that $\rho$ is a density matrix, $\operatorname{tr}(\rho)=1$, Eq. (\ref{fidel}) reduces to:
\begin{eqnarray}
   F(\left\lvert \psi\right\rangle,\rho)&&=\frac{1}{4}\Big(1-\operatorname{tr}(\sigma_z\otimes\sigma_z\rho)+\sin{2\varepsilon_{+-}}\nonumber\\&&[\operatorname{tr}(\sigma_x\otimes\sigma_y\rho)-\operatorname{tr}(\sigma_y\otimes\sigma_x\rho)]+\\&&(1-2\cos^2{\varepsilon_{+-}})[\operatorname{tr}(\sigma_z\otimes\operatorname{Id}\rho)-\operatorname{tr}(\operatorname{Id}\otimes\sigma_z\rho)]\Big).\nonumber\label{eq:fidbs}
\end{eqnarray}

\begin{figure}
    \centering
    \includegraphics[width=0.45\textwidth]{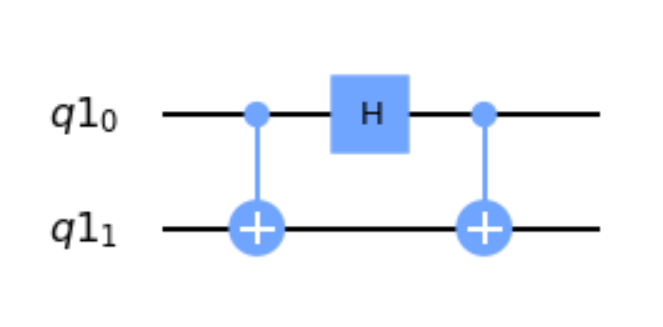}
    \caption{Quantum circuit implementing the equivalent of an X-basis measurement in the beam-splitter case with one maximum excitation per mode, where each mode can be treated as a qubit.}
    \label{fig:XXX} 
\end{figure}

\begin{figure}
    \centering
    \includegraphics[width=0.45\textwidth]{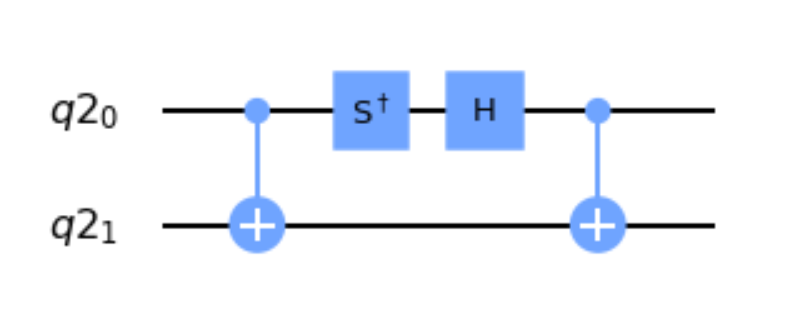}
    \caption{Quantum circuit implementing the equivalent of a Y-basis measurement in the beam-splitter case with one maximum excitation per mode, where each mode can be treated as a qubit.}
    \label{fig:YY} 
\end{figure}

By considering $\operatorname{tr}(\sigma_i\otimes\sigma_j\rho)$ as a mean value of the corresponding observables over $\rho$, we have:
\begin{align*}
    \operatorname{tr}(\sigma_z\otimes\sigma_z\rho)&=p_{ZZ}(\left\lvert 0\right\rangle_+\left\lvert 0\right\rangle_-)+p_{ZZ}(\left\lvert 1\right\rangle_+\left\lvert 1\right\rangle_-)\\&-p_{ZZ}(\left\lvert 0\right\rangle_+\left\lvert 1\right\rangle_-)-p_{ZZ}(\left\lvert 1\right\rangle_+\left\lvert 0\right\rangle_-)\\
    \operatorname{tr}(\sigma_x\otimes\sigma_y\rho)&=p_{XY}(\left\lvert 0\right\rangle_+\left\lvert 0\right\rangle_-)+p_{XY}(\left\lvert 1\right\rangle_+\left\lvert 1\right\rangle_-)\\&-p_{XY}(\left\lvert 0\right\rangle_+\left\lvert 1\right\rangle_-)-p_{XY}(\left\lvert 1\right\rangle_+\left\lvert 0\right\rangle_-)\\
    \operatorname{tr}(\sigma_y\otimes\sigma_x\rho)&=p_{YX}(\left\lvert 0\right\rangle_+\left\lvert 0\right\rangle_-)+p_{YX}(\left\lvert 1\right\rangle_+\left\lvert 1\right\rangle_-)\\&-p_{YX}(\left\lvert 0\right\rangle_+\left\lvert 1\right\rangle_-)-p_{YX}(\left\lvert 1\right\rangle_+\left\lvert 0\right\rangle_-)\\
    \operatorname{tr}(\sigma_Z\otimes\operatorname{Id}\rho)&=p_{Z}(\left\lvert 0\right\rangle_+)-p_{Z}(\left\lvert 1\right\rangle_+)\\
    \operatorname{tr}(\operatorname{Id}\otimes\;\sigma_Z\;\rho)&=p_{Z}(\left\lvert 0\right\rangle_-)-p_{Z}(\left\lvert 1\right\rangle_-).
\end{align*}
The notation $p_{XY}(\left\lvert 0\right\rangle_+\left\lvert 1\right\rangle_-)$ represents the probability of mode $(+)$ being in state 0 of the $X$ base and mode $(-)$ in state 1 of the $Y$ base. Mapping bosons to qubits, the eigenstates of the $X$ and $Y$ bases are:
\begin{equation*}
    \left\lvert0_X\right\rangle_i=\frac{\left\lvert01\right\rangle+\left\lvert10\right\rangle}{\sqrt{2}} \quad\quad\quad
    \left\lvert1_X\right\rangle_i=\frac{\left\lvert01\right\rangle-\left\lvert10\right\rangle}{\sqrt{2}}
\end{equation*}
\begin{equation*}
    \left\lvert0_Y\right\rangle_i=\frac{\left\lvert01\right\rangle+i\left\lvert10\right\rangle}{\sqrt{2}} \quad\quad\quad
    \left\lvert1_Y\right\rangle_i=\frac{\left\lvert01\right\rangle-i\left\lvert10\right\rangle}{\sqrt{2}}
\end{equation*}
where $i=+,-$ labels each bosonic mode. These eigenstates are associated to the measurements that can be seen in Figures \ref{fig:XXX} and \ref{fig:YY}.

Finally, we post-select to normalize the probabilities to the true states of the Hilbert space of the system as explained above.  The fact that in this case we can treat the modes as qubits allows us to realize the post selection in the three basis.  Post-selection probabilities depend strongly on the parameters and the chosen basis, ranging from 40 \% to 80 \%. 

We see in Fig. \ref{fig:pi/36} and \ref{fig:santiago} the whole circuit for $\varepsilon_{+-}=\pi/36$ in Santiago and the final transpiled version, while in Fig. \ref{fig:CASABLANCAA} we present the full circuit in Casablanca for $\varepsilon_{+-}=\pi/2$.

\begin{figure}
    \centering
    \includegraphics[width=0.45\textwidth]{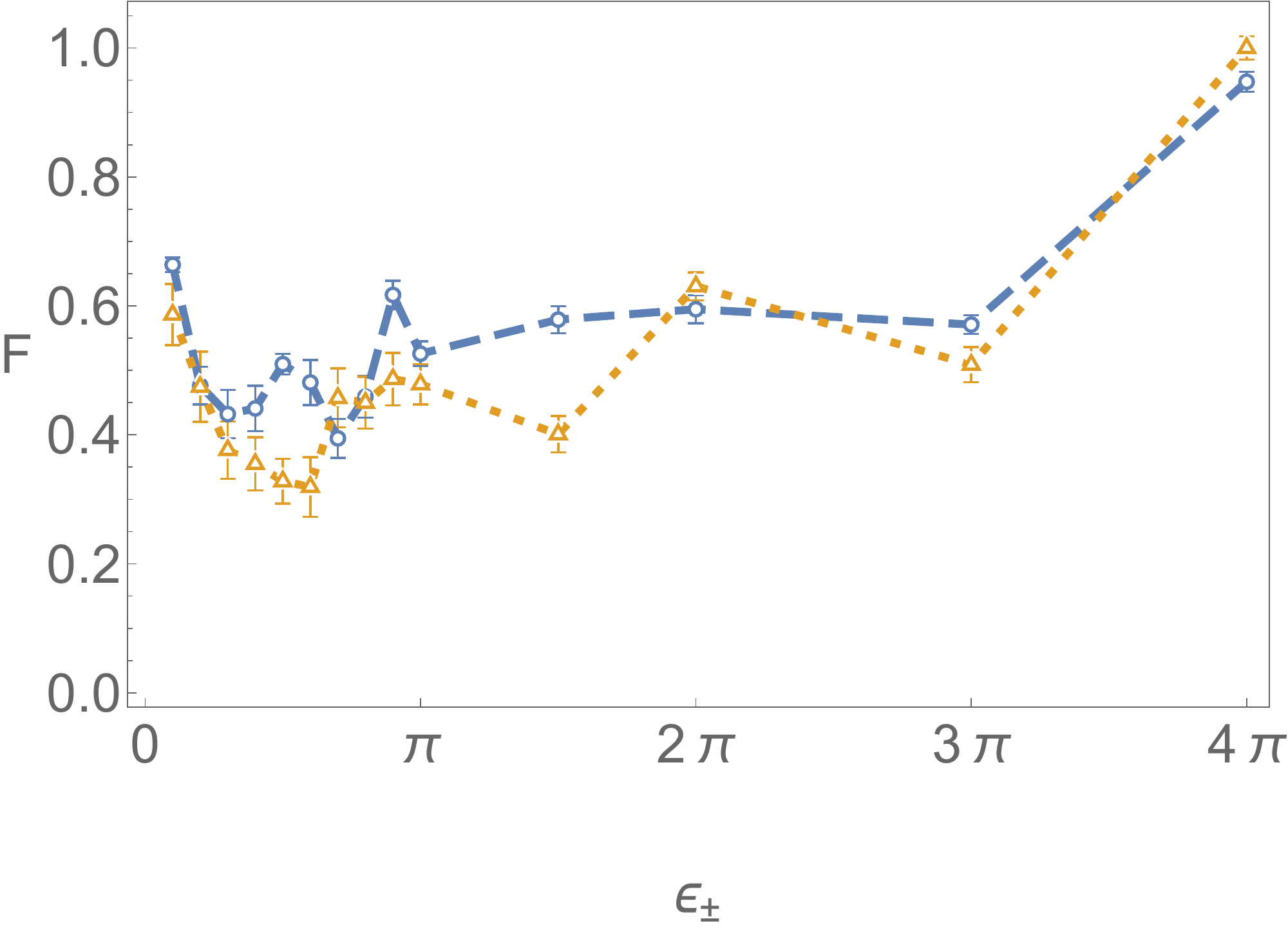}
    \caption{Fidelity $F$ results for the digital quantum simulation of a beam-splitter Eq. (\ref{eq:fidbs}) with squeezing parameter $\varepsilon_{+-}$ in Santiago (blue, dashed, circles) (May/26/2021) and Casablanca (May/24-25/2021) (orange, dotted, triangles).}
    \label{fig:figbs}
\end{figure}

We present the results in Fig. \ref{fig:figbs}. We see a very special value $\epsilon_{+-}=4\pi$, which corresponds to $\frac{\epsilon_{+-}}{8}=\frac{\pi}{2}$ for each of the blocks, reaching an extremely high fidelity. This is due to the fact that in this particular case, the final transpiled circuit contains only 25 CNOT gates and 46 single-qubit gates, meaning that the expected fidelity from error gates is around 85 \%. Moreover, the error mitigation and post-selection work very well and are able to further rise the fidelity.  For other values of $\epsilon_{+-}$, the large number of gates -around 100 CNOT gates and 100 single-qubit gates, typically- generally translates into poor fidelity barely reaching values above 60 \% for some small values of the parameter.

We might try to improve the fidelity by \emph{trotterizing}
our circuit, that is dividing the dynamics into shorter parameter intervals by applying the Trotter formula.
This can be achieved by decomposing each operator $e^{-i\frac{\epsilon_{+-}}{8}\sigma_z}$ into n steps $\prod_{k=1}^n e^{-i\frac{\epsilon_{+-}}{8n}\sigma_z}$ . Note that this will only add extra 1-qubit gates and not 2-qubit ones which are the main source of noise. %which introduce additional noise
To simulate this, we need to take into account that transpiling the circuit gathers all the steps together in a single rotation. So we need to insert barriers between steps to actually achieve \emph{trotterization}.

\begin{figure}
    \centering
    \includegraphics[width=0.45\textwidth]{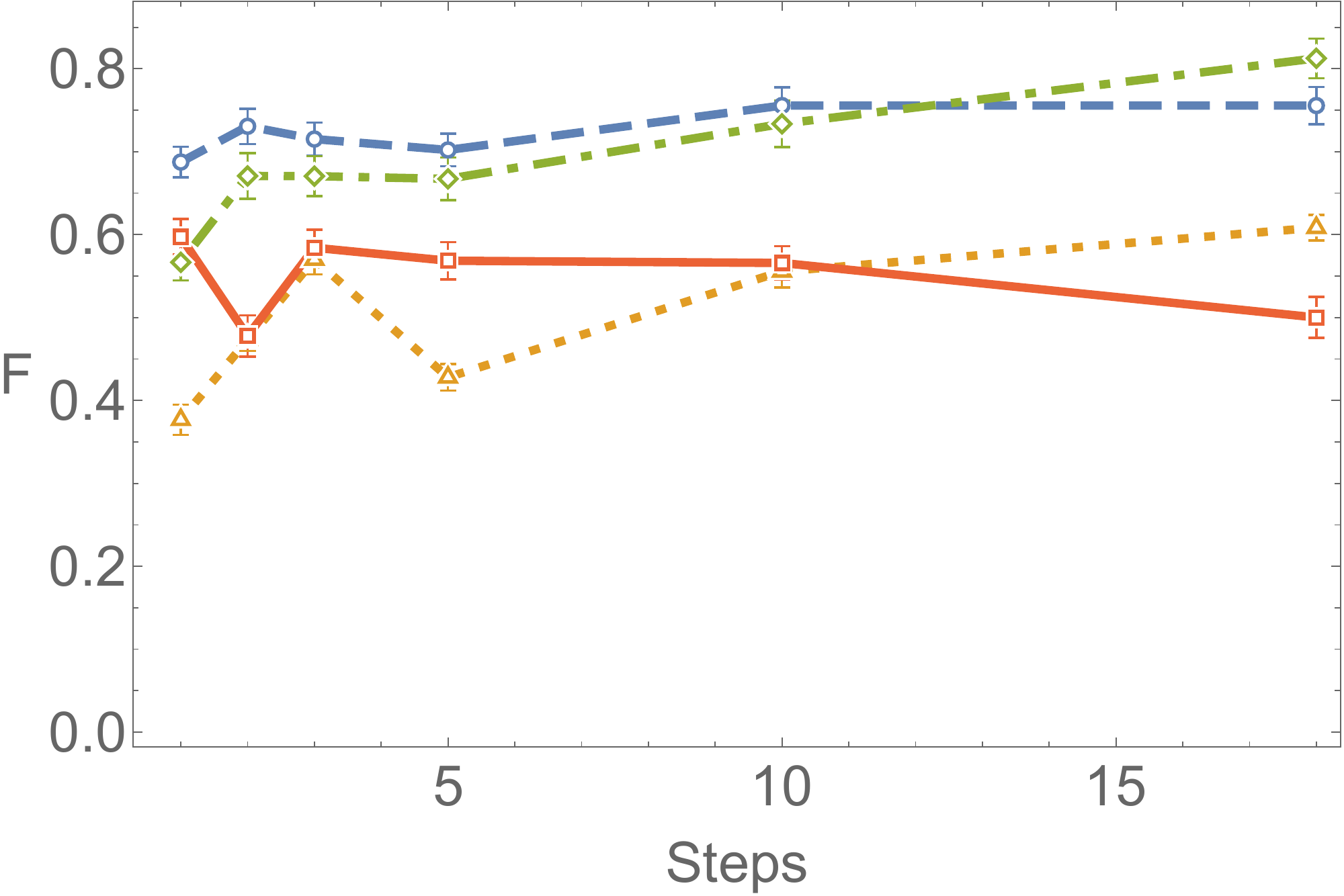}
    \caption{Fidelity $F$ results for the digital quantum simulation of a beam-splitter vs. Trotter steps for Santiago (May/26/2021)  $\varepsilon_{+-}=\pi/36$ (blue, dashed, circles), $\varepsilon_{+-}=\pi/2$ (orange, dotted, triangles) and Casablanca (May/26/2021) $\varepsilon_{+-}=\pi/36$ (green, dash-dotted, diamonds), $\varepsilon_{+-}=\pi/2$ (red, solid, squares).}
    \label{fig:figbstrot}
\end{figure}

We present the fidelities of \emph{trotterization} for $\varepsilon_{+-}=\pi/2$  and $\pi/36$ in Fig. \ref{fig:figbstrot} both in Casablanca and Santiago. For $\varepsilon_{+-}=\pi/36$ we observe a clear improvement after a few steps. As expected, the benefits of trotterization should be suppressed after a certain number of steps due to the noise of the additional barriers. This seems to occur in Santiago after 10 steps, while in Casablanca fidelity keeps moderately improving even after 18 steps. However, for
$\varepsilon_{+-}=\pi/2$ the fidelity gain is only significant after 3 steps in Santiago.

\section{Two-mode squeezing}

\begin{figure}
    \centering
    \includegraphics[width=0.45\textwidth]{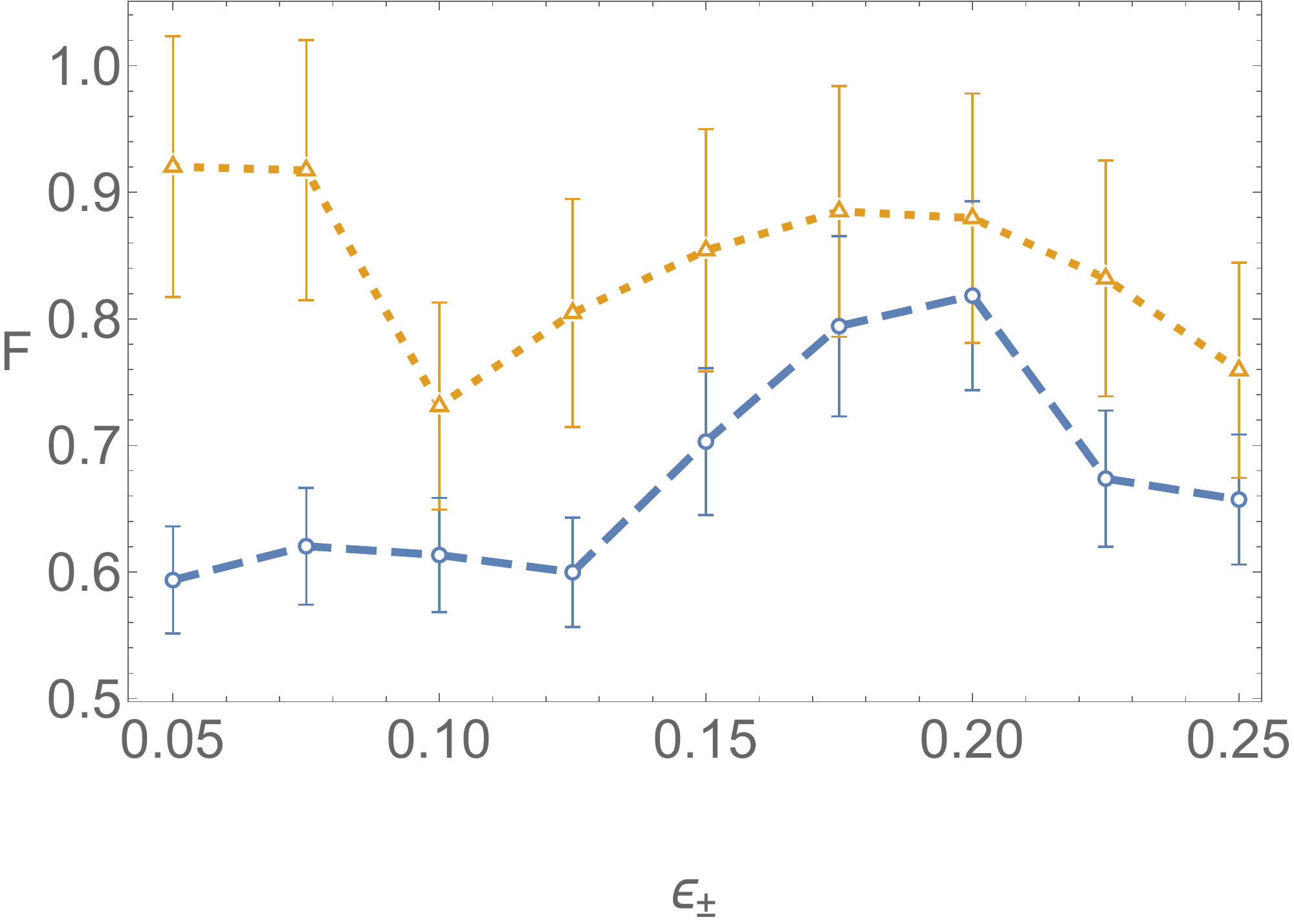}
    \caption{Fidelity results for the digital quantum simulation of two-mode squeezing vs squeezing parameter for Santiago (May/28/2021) (orange, dotted, triangles) and Casablanca (May/29/2021) (blue, dashed, circles).}
    \label{fig:fig2ms}
\end{figure}

\begin{figure*}
    \includegraphics[width=\textwidth]{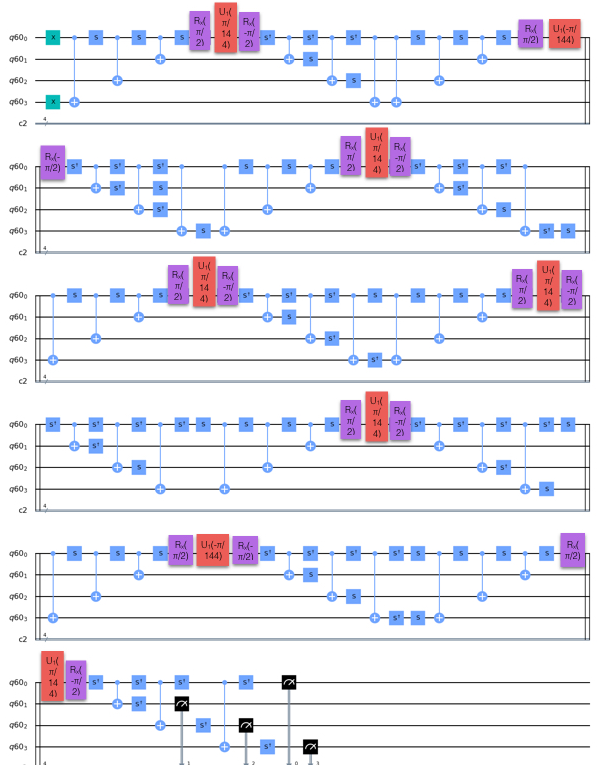}
    \caption{Full circuit for digital quantum simulation of a beam-splitter for $\varepsilon_{+-}=\pi/36$ as launched in Santiago. The order of the qubits is the same as in the text.}
    \label{fig:pi/36} 
\end{figure*}

\begin{figure*}
    \includegraphics[width=0.92\textwidth]{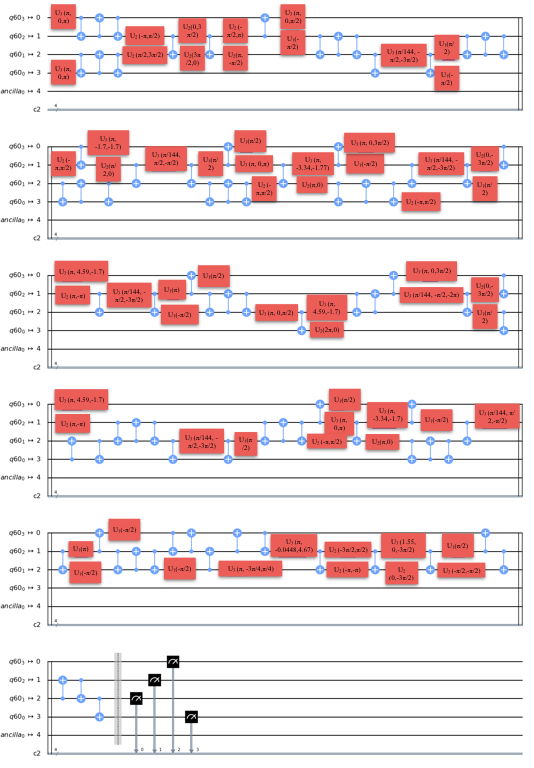}
    \caption{Final transpiled version of the circuit in Figure (\ref{fig:pi/36}).}
    \label{fig:santiago} 
\end{figure*}

\begin{figure*}
    \centering
    \includegraphics[width=\textwidth]{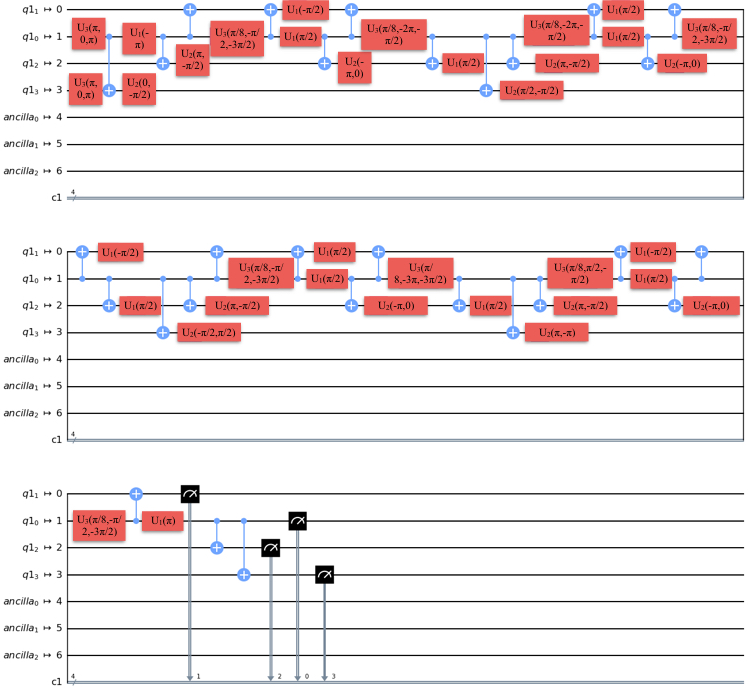}
    \caption{Full circuit for digital quantum simulation of a beam-splitter for $\varepsilon_{+-}=\pi/2$ as launched in Casablanca. $q1_0$, $q1_1$, $q1_2$ and $q1_3$ are respectively qubits 0, 1, 2 and 3 in the text.}
    \label{fig:CASABLANCAA} 
\end{figure*}

\begin{figure*}
    \centering
    \includegraphics[width=\textwidth]{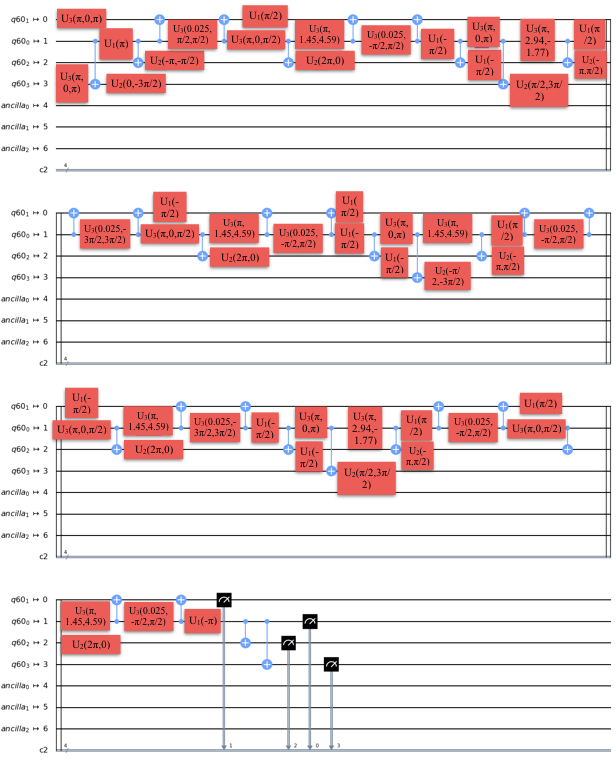} 
    \caption{Full two-mode squeezing transpiled circuit in Casablanca for $\varepsilon=0.1$. $q60_0$, $q60_1$, $q60_2$ and $q60_3$ are respectively qubits 0, 1, 2 and 3 in the text.}
    \label{fig:2 mode} 
\end{figure*}

Finally, as discussed above, we consider the two-mode squeezing interaction allowing only one excitation per mode and starting in the vacuum state. This configuration gives rise to a hamiltonian similar to the beam-splitter for two modes with a maximum of one
photon per mode. They both have a similar series of four-qubit interaction terms, with the only difference of some of the signs. As in the beam-splitter case, we have again two qubits per mode and the same definition for the bosonic operators.
In particular, we consider the unitary
\begin{equation}
    U_{+-}=e^{i\varepsilon_{+-}(b_+^{\dag}a_-^{\dag}+ h.c.)},
\end{equation}
which gives rise to the four-qubit interaction in Eq. (\ref{eq:s4sigmas}).
Using similar techniques as in previous sections, we aim to compute the fidelity after postselection by using the perturbative approximation in Eq. (\ref{eq:fidapprox}). The corresponding transpiled circuit in Santiago is shown in Fig. \ref{fig:2 mode}.
We show the fidelity results in Fig. \ref{fig:fig2ms} for both Casablanca and Santiago, with fidelities ranging from 60 to 90 \%. As in previous cases, both the post-selection probabilities and the final number of gates depend strongly on the parameters, and so does the fidelity.

Let us know briefly discuss the qubit requirements to simulate state-of-the-art quantum-optical experiments. In \cite{photbs}, 25 two-mode squeezed states are generated with squeezing parameters around 1.5, meaning approximately 5 photons per mode. This would amount to approximately 300 qubits in our setup. Then, a linear-optical network of 100 modes with the 25 two-mode squeezed sources above as input is implemented. This would rise the number of qubits up to several thousands. While current superconducting-qubit quantum computers contain less than 100 qubits, IBM plans to reach the threshold of 1000 qubits in 2023 with a 1,121-qubit device called Condor and Google has announced its goal of building a one-million qubit quantum computer by the end of this decade.

\section{Conclusions}

We present results of the digital quantum simulation of quantum-optical single-mode and two-mode bosonic interactions such as beam-splitter and squeezing. We use a boson-qubit mapping in order to translate bosonic hamiltonians into multiqubit gates. Then we apply gate-decomposition techniques to express them as a sequence of single-qubit and CNOT gates to launch the circuits in IBM quantum devices. We make use of the available error mitigation strategies -only in the measurement stage- and also post-selection in order to achieve high-fidelity simulations, which -where possible- we verify by full tomography of the state -otherwise, we use analytical approximations to the fidelity.  We also analyze the usefulness of Trotter techniques. The achieved fidelities are above 90 \% for low single-mode squeezing, diminishing with increasing values of the squeezing parameter. For two-mode interactions such as two-mode squeezing and beam-splitting, the quantum circuits are much more complex and include many more two-qubit gates: we achieve fidelities which range from 60 \% to 90\%, depending on the parameters.

Further generalizations such as higher-order squeezing \cite{threemodes}, higher number of photons in single-mode and two-mode squeezing or sequences of beam-splitters for boson sampling applications will require enhanced connectivity and quantum volume to keep high fidelity.

\section*{Acknowledgements}
P. C. E. have received financial support from CSIC JAE-Intro program (JAEINT-20-02182). A.A. and C.S have received financial support through the Postdoctoral Junior Leader Fellowship Programme from la Caixa Banking Foundation (LCF/BQ/LR18/11640005).



\begin{thebibliography}{25}
\bibitem{gao} Y. Y. Gao, B. J. Lester, Y. Zhang, C. Wang, S. Rosenblum, L. Frunzio, L. Jiang, S. M. Girvin and R. J. Schoelkopf, Phys. Rev. X \textbf{8}, 021073 (2018).
\bibitem{zhanggirvin} Y. Zhang, B. J. Lester, Y.Y. Gao, L. Jiang, R. J. Schoelkopf and S. M. Girvin, Phys. Rev. A \textbf{99}, 012314 (2019).
\bibitem{schoelkopfgirvin} Y. Y. Gao, B. J. Lester, K. S. Chou, L. Frunzio, M. H. Devoret, L. Jiang, S. M. Girvin and R. J. Schoelkopf, Nature \textbf{566} 509 (2019).
\bibitem{pollo} J. Huh, G. G. Guerreschi, B. Peropadre, J. R. McClean and A. Aspuru-Guzik, Nature Photonics \textbf{9}, 615 (2015).
\bibitem{polloexp} C. S. Wang, J. C. Curtis, B. J. Lester, Y. Zhang, Y. Y. Gao, J. Freeze et al. Phys. Rev. X \textbf{10}, 021060 (2020).
\bibitem{supremacybs} H.-S. Zhong, H. Wang, Y.-H. Deng, M.-C. Chen, L.-C. Peng,
Y.-H. Luo et al. Science \textbf{370}, 1460 (2020).
\bibitem{boixo}S. Boixo et al. Nature Physics \textbf{14}, 595 (2018).
\bibitem{naturemonty} A. W. Harrow and A. Montanaro, Nature \textbf{549}, 203 (2017).
\bibitem{arutesup} F. Arute, K. Arya, R. Babbush, D. Bacon, J. C. Bardin, R. Barends et al. Nature \textbf{574} 505 (2019).
\bibitem{sabindqs} C. Sabín, Quantum Reports \textbf{2}, 208 (2020).
\bibitem{lloyd}S. Lloyd, Science \textbf{273}, 1073 (1996).
\bibitem{sommalallama} R. Somma, G. Ortiz, J. E. Gubernatis, E. Knill and R. Laflamme Phys. Rev. A \textbf{65}, 042323 (2002).
\bibitem{losalamos}R. Somma, G. Ortiz, E. Knill and J. Gubernatis, Proc. SPIE 5105, Quantum Information and Computation 96, (2003).
\bibitem{sommathesis} Quantum computation, Complexity and many-body physics, Rolando Somma Phd. Thesis (2005). 
\bibitem{gravwaves} M. Tse et al. Phys. Rev. Lett. \textbf{123} 231107 (2019).
\bibitem{qiskit} Qiskit: An Open-source Framework for Quantum Computing
G. Aleksandrowicz, Online book (2019).
%\bibitem{supp} See Supplemental Material at [URL will be inserted by publisher] for the codes to obtain the results of the manuscript.
\bibitem{nielsenchuang} M. A. Nielsen and I. L. Chuang, \textit{Quantum computation and quantum information}, Cambridge University Press (2010).
\bibitem{mitigation} A. Kandala, K. Temme, A. D. Córcoles, A. Mezzacapo, J. M. Chow and J. M. Gambetta Nature \textbf{567}, 491 (2019).
\bibitem{quantumvolume} Andrew W. Cross, Lev S. Bishop, Sarah Sheldon, Paul D. Nation, and Jay M. Gambetta Phys. Rev. A \textbf{100}, 032328 (2019).
\bibitem{dce} C. M. Wilson, G. Johansson, A. Pourkabirian, M. Simoen, J. R. Johansson, T. Duty, F. Nori and P. Delsing, Nature \textbf{479}, 376-379 (2011).
\bibitem{threemodes} C. W. S. Chang et al. Phys. Rev. X \textbf{10}, 011011 (2020).
\bibitem{photbs} Zhong et al., Science \textbf{370}, 1460–1463 (2020).

\end{thebibliography}
\end{document}